\documentclass[runningheads]{llncs}
\usepackage[T1]{fontenc}
\usepackage{graphicx,verbatim}
\usepackage{fancyhdr}
\usepackage{amsmath} 
\usepackage{array}
\usepackage{graphicx}
 \usepackage{amsmath} 
 \usepackage{booktabs}
\usepackage{multirow}
\usepackage{graphicx}
\usepackage{booktabs} 
\usepackage{subcaption}
\usepackage{amsmath}
\usepackage{xcolor}
\usepackage{multicol}
\usepackage{multirow}
\usepackage{mathrsfs}
\usepackage{amssymb}
\usepackage{xcolor}
\usepackage{textcomp}
\usepackage{pdfpages}
\usepackage{wrapfig}
\usepackage{subcaption}
\usepackage{pdflscape}
\usepackage{pdfpages}
\usepackage{float}
\usepackage{multirow}
\usepackage[nottoc]{tocbibind}
\usepackage{afterpage}
\usepackage{tikz, lipsum,lmodern}
\usepackage{calligra,frcursive,xcolor}
\usepackage{pifont}
\usepackage[most]{tcolorbox}
\usepackage{fancyvrb}
\usepackage{array}
\usepackage{tabularx}
\usepackage{graphicx}  % For \scalebox
\usepackage{booktabs}  % For \midrule, \bottomrule
\usepackage{xcolor}    % For \textcolor
\usepackage{multirow}  % For \multirow
\usepackage{makecell}  % Enables text wrapping inside multirow cells

\usepackage{multirow}
\usepackage{pgfplots}
\usepackage{indentfirst}
\begin{document}
%
% \title{Two-stage Deep Learning Network for Atrial Structure Segmentation}

\title{TASSNet: A Deep Learning Framework for Robust Bi-Atrial Segmentation for Assessing Structural Basis of Atrial Fibrillation}

\author{Malitha Gunawardhana \and
Mark L Trew \and
Gregory B Sands \and
Jichao Zhao}
\authorrunning{Gunawardhana et al.}

\institute{Auckland Bioengineering Institute, University of Auckland, New Zealand
\email{malitha.gunawardhana@auckland.ac.nz}}

\maketitle              

\begin{abstract}

Atrial Fibrillation (AF), the most common sustained cardiac arrhythmia worldwide, increasingly requires accurate bi-atrial structural assessment to guide ablation strategies, particularly in persistent AF. 
Late gadolinium-enhanced magnetic resonance imaging (LGE-MRI) enables visualisation of atrial fibrosis, but precise manual segmentation remains time-consuming, operator-dependent, and prone to variability. 
We propose TASSNet, a novel two-stage deep learning framework for fully automated segmentation of both left atrium (LA) and right atrium (RA), including atrial walls and cavities, from 3D LGE-MRI. TASSNet introduces two main innovations: (i) a ResNext-based encoder to enhance feature extraction from limited medical datasets, and (ii) a cyclical learning rate schedule to address convergence instability in highly imbalanced, small-batch 3D segmentation tasks. We evaluated our method on two datasets, one of which was completely out-of-distribution, without any additional training. In both cases, TASSNet successfully segmented atrial structures with high accuracy. These results highlight TASSNet’s potential for robust and reproducible bi-atrial segmentation, enabling advanced fibrosis quantification and personalised ablation planning in clinical AF management.

\end{abstract}

\section{Introduction}

Atrial fibrillation (AF) is the most common sustained cardiac arrhythmia~\cite{lippi2021global}, with growing evidence highlighting its bi-atrial nature, particularly in persistent AF, where both the left atrium (LA) and right atrium (RA) are involved~\cite{bax2022effect,kottkamp2013human}. While pulmonary vein isolation (PVI) remains standard therapy~\cite{oral2022}, its limited efficacy in persistent AF underscores the need for more comprehensive ablation strategies targeting fibrotic substrates~\cite{oral2022,kistler2023effect,roney2020silico}.

Late gadolinium-enhanced MRI (LGE-MRI) enables non-invasive visualisation of atrial fibrosis and has shown potential in guiding patient-specific ablation, as demonstrated in studies such as DECAAF II\cite{marrouche2021efficacy}. However, precise segmentation of atrial structures from LGE-MRI remains challenging due to thin atrial walls, complex anatomy, and substantial inter-observer variability~\cite{higuchi2018spatial} while manual segmentation is laborious, error-prone and resource-intensive. 

With the rise of deep learning, automated segmentation of cardiac MRI has emerged as a promising solution for addressing these challenges. CNNs have proven effective for image analysis tasks, particularly in facilitating automatic segmentation of various cardiac structures from LGE-MRIs~\cite{xu2024dynamic,gunawardhana2024good}. However, despite these advancements, most studies remain limited by their small, single-centre datasets, reducing the generalizability of the models\cite{zhuang2019evaluation}. Additionally, these studies often focus solely on the LA, neglecting the RA in the process, even though it has been recognised that AFib is a bi-atrial disease~\cite{bax2022effect,kottkamp2013human}.

In this work, we propose TASSNet, a two-stage deep learning framework for fully automated bi-atrial segmentation from 3D LGE-MRI. TASSNet uses ResNext-based encoders~\cite{xie2017aggregated}, which leverage grouped convolutions to capture richer feature representations with fewer parameters, improving generalisation in small medical datasets where high anatomical variability exists. The two-stage design first localises atrial regions using coarse 3D segmentation, followed by fine segmentation within the extracted regions using combined 2D and 3D networks. A cyclical learning rate schedule further stabilises training in this highly imbalanced 3D setting.

We comprehensively benchmark TASSNet against several state-of-the-art models using two datasets, one of which is out-of-distribution (OOD), without any additional training. This work presents a robust framework for bi-atrial segmentation, representing a critical step toward fibrosis quantification and personalised AF therapy.

\section{Methodology}

\begin{figure}[!h]
\centering
\includegraphics[width=\linewidth]{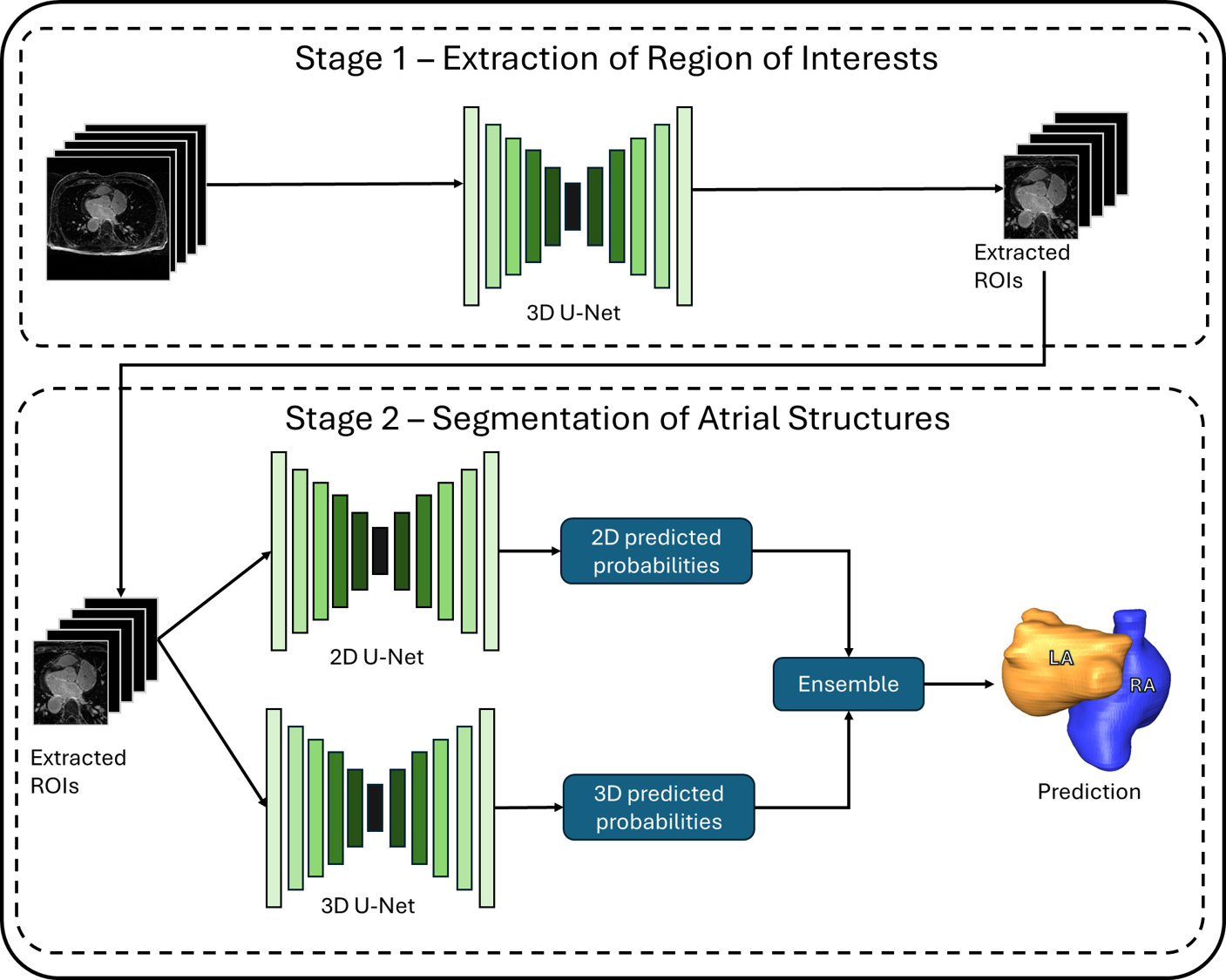}
\caption{Overview of the TASSNet framework.  In Stage 1, a 3D U-Net is employed to extract regions of interest (ROIs) from the input LGE-MRI volume, concentrating on the atrial structures. The output of stage I is a localised ROI, reducing spatial complexity for more precise segmentation in stage 2. In Stage 2, both 2D and 3D U-Net architectures are used to segment the atrial structures within the extracted ROIs finely. The 2D U-Net processes data slice-by-slice, generating 2D probability maps, while the 3D U-Net uses volumetric information to produce 3D probability maps. The predicted 3D model, delineating the left and right atrial walls and cavities, is the ensemble output of both networks.}
\label{fig:archi}
\end{figure}

This study introduces a two-stage deep learning framework called TASSNet aimed at segmenting atrial structures from 3D LGE-MRI data, which is particularly challenging due to the complexity of atrial anatomy and the inherent class imbalance in medical imaging datasets. As shown in Fig. \ref{fig:archi}, the framework strategically decomposes the task into two key stages: ROI extraction and refined segmentation.

\subsection{Stage 1: Extraction of Regions of Interest}

The first stage leverages a 3D U-Net~\cite{ronneberger2015u} architecture to perform coarse segmentation of atrial structures. This coarse segmentation serves as a preliminary localisation step designed to identify the general regions of interest (ROI) within the 3d LGE-MRI volumes that encompass the atria. By focusing only on these regions, the model effectively reduces the computational load and addresses the prevalent issue of class imbalance, which arises from the disproportionate size of the atria relative to the entire MRI volume. 

Following this coarse segmentation, the centre of mass of the segmented atria is computed, which enables the extraction of a fixed-size patch around the atrial region. A patch size of 256x256x44 ensures that the extracted region covers the atria, discarding the background. This significantly reduced the complexity of the subsequent fine segmentation task as the network can focus on a simpler and more relevant subset of the image.

\subsection{Stage 2: Segmentation of Atrial Structures}

Once the ROIs are extracted, the second stage focuses on the refined segmentation of the atrial structures. In this stage, two separate U-Nets are employed, one operating on 2D slices and the other on 3D volumes of the extracted ROIs. This helps to learn a richer spatial context during the segmentation. 

Both networks predict the likelihood of each voxel or pixel (in the case of 2D) belonging to a specific class. These predictions, represented as probability maps, are then ensembled to produce the final segmentation mask. This ensembling approach capitalises on the complementary strengths of both 2D and 3D networks. The 2D network excels in capturing fine details on a slice-by-slice basis, while the 3D network is more adept at maintaining spatial coherence across multiple slices, making the combination of their outputs more robust and accurate. Post-processing restores the predicted segmentation masks to the original size of the 3D LGE-MRI, ensuring correct spatial alignment and anatomical accuracy.

\subsection{Network Architecture}

The U-Net architecture used in both stages is based on a modified version of the classical U-Net, enhanced with ResNext \cite{xie2017aggregated} blocks and instance normalisation (InstanceNorm) \cite{ulyanov2016instance} to improve convergence during training. The decision to employ ResNext blocks instead of the standard block or ResNet \cite{he2016deep} blocks was motivated by the superior ability of ResNext to aggregate multiple transformations efficiently, leading to better feature representation and improved performance, especially in high-dimensional medical imaging data. 

The architecture consists of seven stages, with each stage employing convolutional layers with 3x3x3 kernels in the 3D U-Net (and 3x3 kernels in the 2D U-Net). The number of features in the network starts at 32 in the first layer and progressively increases to 64, 128, 256, and eventually 512 in the deeper layers of the network. The final three stages of the encoder maintain a consistent 512 features, ensuring that high-level abstract representations of the input data are captured effectively. This deep, progressive feature extraction allows the network to handle the inherent variability in the shapes and sizes of the atrial structures across different images while also preserving fine-grained details critical for accurate segmentation. The incorporation of ResNext blocks further enhances the model’s ability to capture both global and local contextual information, making it well-suited for the complex task of atrial segmentation in medical images.

In the proposed training approach, a cyclical learning rate schedule with exponential decay is used. Let \( R \) represent the total number of epochs and \( Z \) the number of learning rate cycles. The number of epochs per cycle  \( T_c = {R}/{Z} \). Within each cycle, the learning rate varies between a maximum value \( lr_r \) and a minimum value \( lr_0 \). The scaling factor \( \beta = {M}/{T_c} \)  controls the exponential decrease of the learning rate.  The learning rate \( lr(i) \) at any epoch \( i \) is defined as follows:

\begin{equation}
lr(i) = 
\begin{cases} 
lr_r & \text{if } t_c = 0 \\
lr_0 + (lr_r - lr_0) \times e^{-\beta \cdot t_c} & \text{if } t_c > 0 
\end{cases}
\end{equation}
where \( t_c = i \mod T_c \) represents the current epoch within the cycle. At the start of each cycle ($t_c = 0$), the learning rate resets to its maximum value $lr_r$, encouraging exploration and helping the model escape local minima. As the cycle progresses ($t_c > 0$), the learning rate decays exponentially from $lr_r$ to $lr_0$, with the decay rate controlled by $\beta$, enabling a smooth transition from exploration to exploitation.

\section{Experiments}
\subsection{Dataset}

This study utilized an LGE-MRI dataset from the University of Utah, comprising 100 3D scans from 41 patients, acquired using 1.5T Avanto or 3.0T Verio scanners~\cite{marrouche2014association,higuchi2018spatial}. The images have an isotropic in-plane resolution of 0.625 × 0.625 mm, a slice thickness of 2.5 mm, and consist of 44 axial slices with in-plane dimensions ranging from 576 × 576 to 640 × 640 pixels.

Manual segmentations of the LA and RA cavities and bi-atrial walls were performed for each scan. Annotations were conducted by three trained observers with over three years of experience in cardiac MRI interpretation and atrial anatomy. A single consensus segmentation was generated per scan, with accuracy ensured through expert review and resolution of any disagreements via consensus discussions. The cavities were defined by manually tracing the blood pools of the LA and RA. For the LA, the segmentation included all four pulmonary veins (PVs), constrained to the antrum region—extending up to 10 mm from the endocardial surface or approximately three times the LA wall thickness—where PV tapering ceased. The mitral and tricuspid valves, connecting the atria to the respective ventricles, were represented using 3D planes to create smooth surfaces separating the atrial and ventricular chambers. The cavity segmentations were morphologically dilated and manually refined to delineate the outer atrial walls. Additional manual editing was performed around complex anatomical regions, such as PVs and valves, and the interatrial septum was traced to connect the LA and RA walls. LA segmentations were provided by the University of Utah~\cite{marrouche2014association}, while RA segmentations were manually generated by the authors following the same protocol.

The dataset was divided into 80 training and 20 test images, ensuring no patient overlap between the sets to prevent data leakage. An ensemble model trained on this dataset was further evaluated on an independent, privately held out-of-distribution (OOD) dataset acquired using the same imaging protocol. As ground truth annotations were unavailable for this dataset, assessment was limited to qualitative visual inspection of segmentation outputs.

\subsection{Implementation Details}

The model was implemented in PyTorch 2.0.1 with a batch size of 4. Coarse segmentation (background vs. ROIs) was trained for 250 epochs, followed by fine segmentation for 1000 epochs with 250 iterations per epoch. Early stopping was used, triggered after 50 and 100 epochs without validation loss improvement in the coarse and fine stages, respectively. Weights were randomly initialised. 

A cyclical learning rate schedule ($Z = 4$) was employed, varying between 0.01 and 0.1 with a scaling factor of $M = 4$. The AdamW optimizer (weight decay 0.01, $\beta_1 = 0.9$, $\beta_2 = 0.999$) was used alongside DiceFocal loss~\cite{ma2021loss} to mitigate class imbalance. Model training was conducted on a Tesla V100 GPU (32 GB).

Extensive online data augmentation, such as rotation, scaling, Gaussian noise, contrast adjustment, gamma correction, inversion, and mirroring, was applied to improve generalisation. The model used a cardinality of $C = 8$, and results were validated with five-fold patient-level cross-validation.

\subsection{Evaluation Metrics}

The effectiveness of the segmentation model is evaluated using three widely adopted metrics: Dice Similarity Coefficient (DSC), Average Surface Distance (ASD), and the 95th percentile of the Hausdorff Distance (HD95). These metrics collectively provide a comprehensive assessment of the overlap between the predicted and ground truth segmentations, as well as the geometric proximity of their respective surfaces.

\begin{table}[!h]
\centering
\setlength{\tabcolsep}{3pt}
\caption[Performance of different methods on the Utah dataset]{Performance of different methods on the Utah dataset for Right Atrium (RA) wall, Left Atrium (LA) wall, RA cavity, and LA cavity. DSC: Dice Similarity Coefficient, ASD: Average Surface Distance (mm), HD95: 95th percentile of Hausdorff Distance (mm). The best values are \textcolor{red}{highlighted}. 3D low.: 3D low resolution, 3D cas. :3D cascade model, Ens.: Ensemble. }
\label{tab:utah_dataset}
\scalebox{0.7}{
\begin{tabular}{p{2.2cm}!{\vrule width 2pt}l!{\vrule width 2pt}l|l|l!{\vrule width 2pt}l|l|l!{\vrule width 2pt}l|l|l!{\vrule width 2pt}l|l|l}

 \toprule 
\multirow{2}{*}{Model}&\multirow{2}{*}{Config}& \multicolumn{3}{c!{\vrule width 2pt}}{RA Wall}& \multicolumn{3}{c!{\vrule width 2pt}}{LA wall}& \multicolumn{3}{c!{\vrule width 2pt}}{RA cavity}& \multicolumn{3}{c}{LA cavity}\\ && DSC& ASD& HD95& DSC& ASD& HD95& DSC& ASD& HD95& DSC& ASD&HD95\\ 
\midrule

\multirow{5}{*}{nnUNet}
&2D & 0.738 & 0.548 & 2.094 & 0.594 & 0.771 & 3.125 & 0.915 & 0.769 & 3.031 & 0.918 & 0.773 & 3.098 \\ 
&3D & 0.735 & 0.557 & 2.069 & 0.609 & {0.739} & \textcolor{red}{2.778} & 0.913 & 0.771 & 2.786 & 0.920 & 0.758 & 2.866 \\ 
&3D low& 0.723 & 0.622 & 2.408 & 0.574 & 0.973 & 4.231 & 0.908 & 0.841 & 3.094 & 0.917 & 0.821 & 3.216 \\ 
&3D cas. & 0.735 & 0.556 & 1.981 & 0.609 & {0.739} & 2.797 & 0.914 & 0.779 & 2.751 & 0.920 & 0.758 & 2.916 \\ 
&Ens. & 0.740 & 0.541& 2.040 & 0.610 & 0.751 & 2.856 & 0.915 & 0.762 & 2.817 & 0.921 & 0.750 & 2.949 \\

\midrule
\multirow{5}{*}{\makecell{nnUNet \\ with ResEnc}} &2D 
& 0.734& 0.551& 2.076& 0.592& 0.741& 2.857& 0.915& 0.741& 2.860& 0.919& 0.756& 2.869\\ 
&3D & 0.737& 0.531& \textcolor{red}{1.920} & 0.606&0.744& 2.819& 0.913& 0.765& 2.812& 0.921& 0.741& 2.776\\ 
&3D (low)& 0.712& 0.656& 2.603& 0.555& 1.016& 4.378& 0.906& 0.862& 3.163& 0.916& 0.817& 3.050\\ 
&3D (cas.)& 0.736& 0.548& 1.952& 0.600& 0.760& 2.881& 0.913& 0.760& 2.765& 0.92& 0.754& 2.808\\ 
 &Ens. & 0.740& 0.538& 1.933&0.605& 0.748& 2.815& 0.915& 0.751& 2.739& 0.922& 0.739& 2.835\\

\midrule
\multirow{3}{*}{UMambaBot}& 2D& 0.730& 0.547& 2.039& 0.598& 0.733& 2.913& 0.912& 0.789& 3.045& 0.918& 0.765&2.905
\\
 & 3D & 0.735& 0.548& 2.087& 0.612& 0.725& 2.814& 0.912& 0.765& 2.738& 0.921& 0.731&2.733
\\
 & Ens.& 0.743& \textcolor{red}{0.514}& 1.939& 0.613& 0.703& 2.771& 0.918& 0.724& 2.667& 0.923& 0.705&2.709\\

 \midrule
 \multirow{3}{*}{UMambaEnc}& 2D
& 0.730& 0.561& 2.137& 0.601& 0.709& 2.799& 0.908& 0.869& 3.482& 0.920& 0.733&2.859
\\
 & 3D 
& 0.737& 0.556& 2.107& 0.617& 0.729& 2.743& 0.913& 0.768& 2.802& 0.921& 0.742&2.777
\\
 & Ens.& 0.746& 0.522& 1.945& 0.618& \textcolor{red}{0.696}& 2.771& 0.916& 0.762& 2.897& 0.923& \textcolor{red}{0.685}&\textcolor{red}{2.620}\\ 

\midrule
\multirow{3}{*}{\makecell{TASSNet \\ with ResNext}}
&2D & 0.742& 0.562& 2.225& 0.608& 0.757& 3.073& 0.915& 0.756& 3.017& 0.918& 0.772& 2.914
\\ 
 &3D & 0.739& 0.641& 2.350& 0.617& 0.838& 3.421& 0.913& 0.815& 3.093& 0.919& 0.812& 3.232
\\ 
 &Ens. & \textcolor{red}{0.753}& 0.564& 2.159& \textcolor{red}{0.620}& 0.739& 3.080& \textcolor{red}{0.921}& \textcolor{red}{0.713}& \textcolor{red}{2.655}& \textcolor{red}{0.924}& {0.702}& {2.684}\\ 
 \bottomrule
\end{tabular}}

\end{table}

\begin{figure}[!h]
    \centering
    \includegraphics[width=\linewidth]{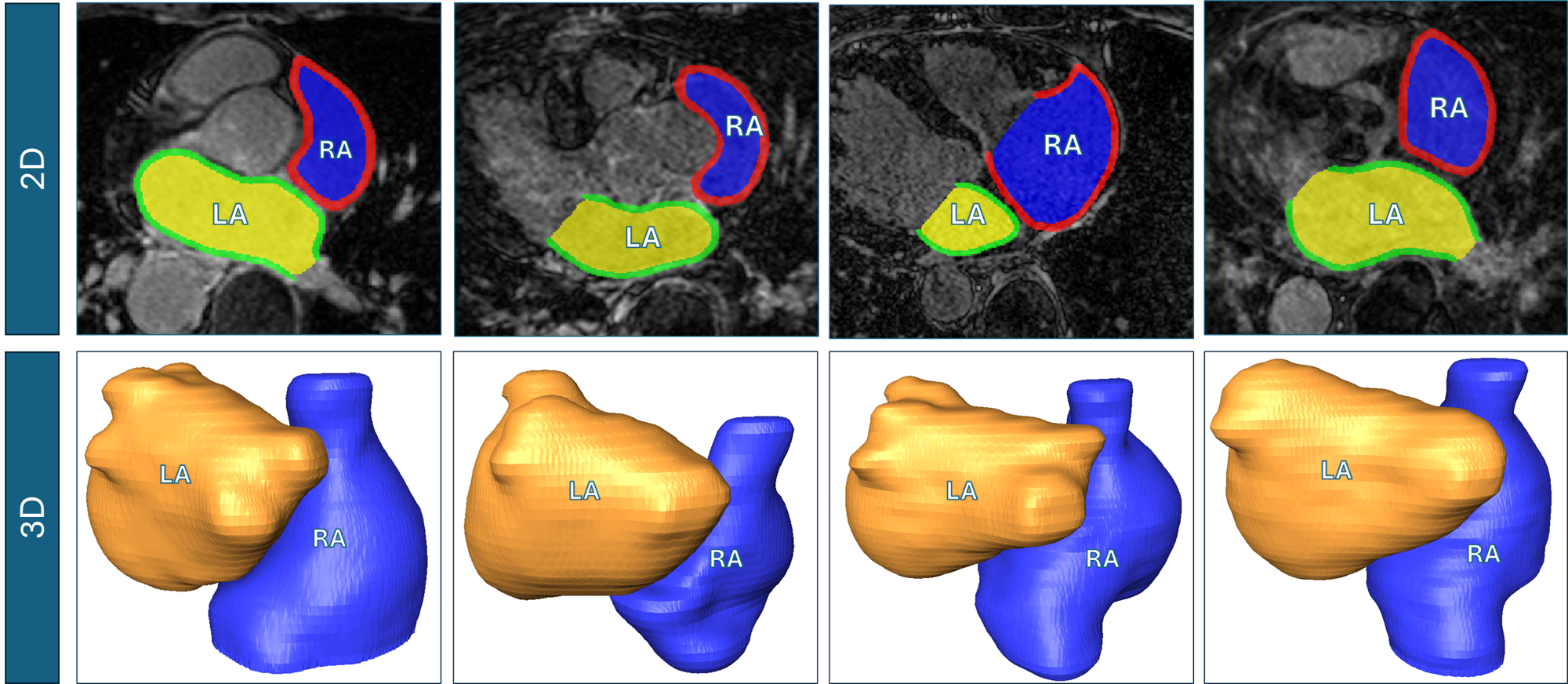}
    \caption{Performance visualisation of the ensemble model in 2D (the first row) and 3D views (the second row) for four different hearts in the Utah dataset. The left atrial (LA) cavity is highlighted in yellow, the right atrial (RA) cavity in blue, the LA wall in green, and the RA wall in red.}
    \label{fig:utah}
\end{figure}

\section{Results}

The performance of the proposed method was compared against several existing segmentation models, including the original nnU-Net framework~\cite{isensee2021nnu,gunawardhana2024good}, nnU-Net variants with ResNet encoders, and two recent Mamba-based architectures: U-MambaBot~\cite{ma2024u} and U-MambaEnc~\cite{ma2024u}. For the nnU-Net baseline, ensemble predictions were generated by combining outputs from four configurations: 2D, 3D full resolution, 3D low resolution, and 3D cascade models. In the U-MambaBot architecture, a Mamba-based bottleneck module is integrated between the encoder and decoder, whereas U-MambaEnc replaces the entire encoder with Mamba layers.

We evaluated the model using three approaches: 2D, 3D, and an ensemble of the two. For the ensemble, we averaged the softmax probability maps produced by the 2D and 3D models. The mean inference time was 4.46 ± 0.71 seconds for 2D and 8.65 ± 1.45 seconds for 3D per scan.

Table~\ref{tab:utah_dataset} summarises the quantitative performance of these models on the Utah dataset across four anatomical structures: the RA wall, LA wall, RA cavity, and LA cavity. Among all models, the TASSNet architecture equipped with ResNext encoders consistently achieved superior performance in terms of Dice scores across all anatomical structures, demonstrating its robust capacity for accurate segmentation. However, for certain metrics such as ASD and HD95 related to wall segmentation, U-MambaBot, U-MambaEnc, and nnU-Net with ResNet encoders demonstrated slightly improved boundary accuracy, highlighting their ability to generate more precise anatomical contours. 

To visually assess model performance, Figure~\ref{fig:utah} depicts representative qualitative results from the ensemble model on four subjects from the Utah dataset. Both 2D and 3D reconstructions are shown for comprehensive visual evaluation. In these visualisations, the LA cavity is highlighted in yellow, the RA cavity in blue, the LA wall in green, and the RA wall in red. The 3D surface reconstructions further illustrate the anatomical plausibility and clinical relevance of the generated segmentations.

\begin{figure}[!h]
    \centering
    \includegraphics[width=\linewidth]{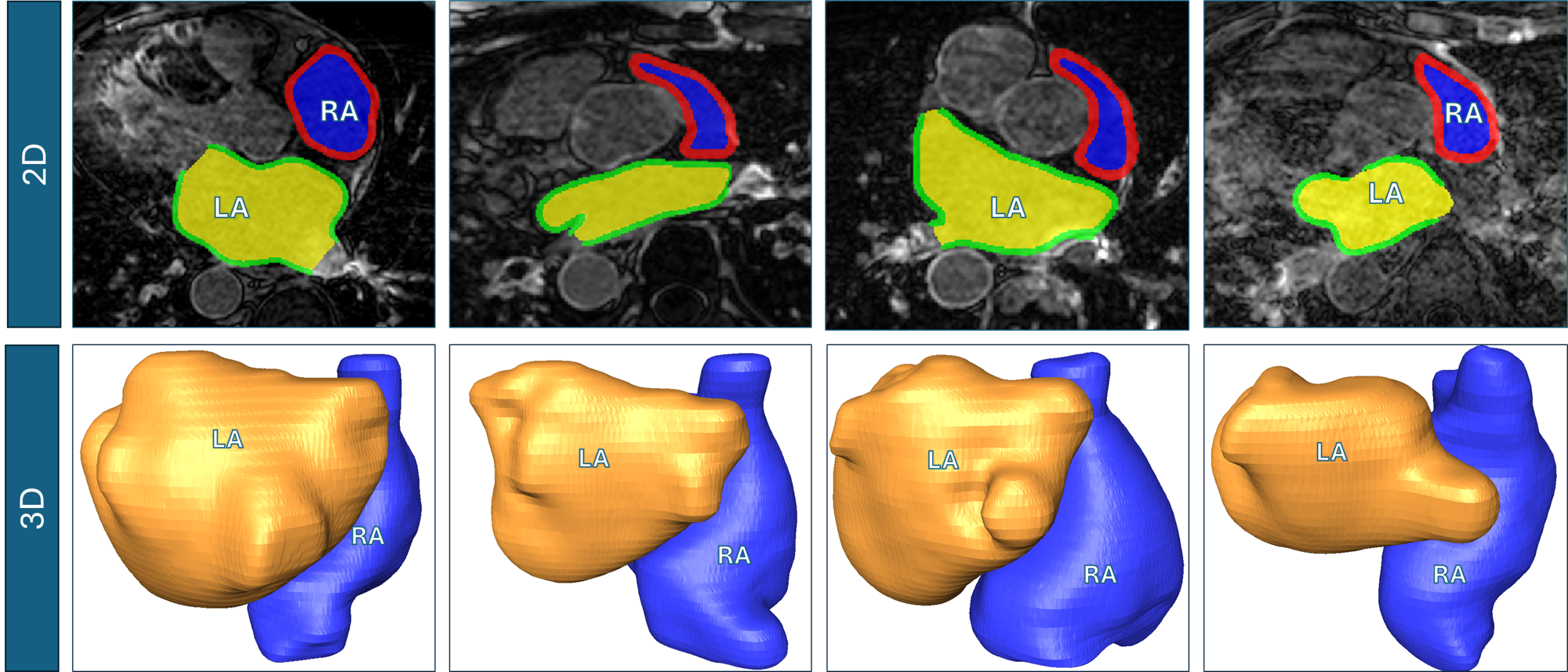}
    \caption{  Performance visualisation of the ensemble model in 2D (the first row) and 3D views (the second row) for four different hearts in the OOD dataset. The left atrial (LA) cavity is highlighted in yellow, the right atrial (RA) cavity in blue, the LA wall in green, and the RA wall in red. }
 
    \label{fig:waikato}
\end{figure}

In Figure~\ref{fig:waikato}, we present 2D and 3D views of four subjects from the OOD dataset. Despite not undergoing any fine-tuning on this new dataset, the TASSNet successfully delineated the anatomical structures, accurately identifying both atrial walls and cavities. These results demonstrate the model's capacity for generalisation and its potential applicability to unseen datasets.

\section{Discussion and Conclusion}

The novel TASSNet framework enables robust and fully automated bi-atrial segmentation from 3D LGE-MRI images, and this opens the potential for using this imaging more widely in AF management especially in measuring clinical biomarkers such as fibrosis, atrial volume and diamter\cite{feng2025automatic}. By accurately segmenting both the LA and RA walls and cavities, the capabilities of TASSNet reflect the emerging consensus that AF is a bi-atrial disease, advancing beyond most existing approaches that focus predominantly on the LA. TASSNet enables comprehensive atrial characterisation and supports the development of more precise, patient-specific therapies.

The grouped convolutions of ResNext enhance feature representation while maintaining parameter efficiency, an important advantage when dealing with relatively small medical datasets and the high inter-patient anatomical variability inherent to AF populations. Further contributing to its stability, TASSNet employs Instance Normalisation in place of BatchNorm, directly addressing convergence issues associated with small 3D batch sizes common in medical imaging applications. 

Unlike prior studies that often evaluated models in isolated or narrowly defined settings~\cite{feng2025automatic}, this work provides a comprehensive benchmarking of multiple architectures under a unified framework, offering a more robust and directly comparable reference point for future research. While TASSNet achieves strong overall performance, accurately delineating thin atrial walls remains a persistent challenge, as reflected in the surface distance metrics. This difficulty highlights a fundamental limitation of LGE-MRI in resolving very thin myocardial structures, suggesting that further algorithmic refinements and potentially multi-modal data fusion may be needed to fully address this issue. Encouragingly, TASSNet demonstrated promising performance even on an OOD dataset, indicating strong potential for generalizability across diverse clinical settings.

Looking forward, the segmentation capabilities of TASSNet lay a solid foundation for downstream tasks such as fibrosis quantification~\cite{tzeis2019atrial,cunha2022atrium}, which is essential for effective risk stratification, personalised treatment planning, and long-term clinical outcome prediction~\cite{akoum2012atrial,li2024research}. Integrating fibrosis maps, fat distribution, and procedural outcome images into the derived 3D model will be essential for transforming imaging advancements into practical, patient-specific clinical decision support tools.

In summary, TASSNet represents a step forward in the automated, reproducible, and clinically relevant segmentation of both atria from LGE-MRI. By addressing key challenges in multi-structure segmentation, data imbalance, and architectural robustness, it provides a scalable platform for integrating advanced imaging biomarkers into precision AF care, ultimately supporting improved patient stratification, individualised therapy guidance, and better long-term outcomes.

\bibliographystyle{splncs04}
\bibliography{mybibliography}

\end{document}